\documentclass[preprint,12pt]{elsarticle}

\usepackage{amssymb}
 \usepackage{amsthm}

\journal{ Physics Letters B}

\begin{document}

\begin{frontmatter}



\title{Dual Doubled Geometry}


\author[ea]{Eric A. Bergshoeff}

\ead{E.A.Bergshoeff@rug.nl}

\author[fr]{Fabio Riccioni}

\ead{fabio.riccioni@roma1.infn.it}

\address[ea]{Centre for Theoretical Physics, University of Groningen, \\ Nijenborgh 4, 9747 AG Groningen, The
Netherlands}

\address[fr]{INFN Sezione di Roma, \\  Dipartimento di Fisica, Universit\`a di Roma ``La Sapienza'' \\ Piazzale Aldo Moro 2, 00185 Roma, Italy}

\begin{abstract}
%It is well-known that a T-duality covariant formulation of the fundamental branes of toroidally compactified string theory
% with 32 supercharges requires a doubled geometry.
We  probe  doubled geometry with dual fundamental branes, i.e.~solitons. Restricting ourselves first to solitonic branes with more than two transverse directions we find that the doubled geometry requires an effective wrapping
rule for the solitonic branes which is dual to the wrapping rule for fundamental branes. This dual wrapping rule can be understood by the presence of Kaluza-Klein monopoles.
Extending our analysis to supersymmetric
solitonic branes with less than or equal to two transverse directions we show that such solitons are
precisely obtained by applying the same dual wrapping rule to these cases as well. This extended wrapping rule can not be
explained by the standard Kaluza-Klein monopole alone. Instead, it
suggests the existence of a class of generalized Kaluza-Klein monopoles in ten-dimensional string theory.
\end{abstract}


\end{frontmatter}

% \linenumbers

\section{Introduction}
\label{introduction}
Over the course of years supergravity has provided  a number of key insights into string theory. Examples of such discoveries are
the Green-Schwarz anomaly cancellation \cite{Green:1984sg} and the presence of branes in string theory such as the eleven-dimensional supermembrane
\cite{Bergshoeff:1987cm}. A key signature for a $p$-brane, i.e.~a brane with $p$ spatial directions, in string theory is the presence of a
$(p+1)$-form potential in the corresponding supergravity theory. Usually, branes have more than two transverse directions and such branes couple to potentials that describe physical degrees of freedom. These are the so-called ``standard'' branes and they are well understood. Whenever a brane
is standard the dual brane, i.e.~the brane that couples to the dual potential, is also standard. The remaining ``non-standard'' branes
are the branes with two transverse directions (``defect''-branes), one transverse direction (``domain-walls'') and no transverse direction at all (``space-filling branes'').\,\footnote{It is well-known that these non-standard branes are not well-defined when considered as single branes.
We will not discuss these issues here and instead only consider whether or not the single brane case is consistent with basic
requirents such as gauge invariance and supersymmetry.}
The special thing about the defect-branes is that they couple to the duals of scalars parametrizing
a coset manifold $G/H$. This leads to $(D-2)$-form potentials transforming in the adjoint of $G$ whose curvatures satisfy dim\,$H$
non-linear constraints that involve the scalars themselves. Due to this there is no one-to-one correspondence between a potential
and a corresponding defect-brane. Domain-walls and space-filling branes are special in the sense that they couple to potentials that do not describe any physical degrees of freedom. In the case of domain walls the corresponding $(D-1)$-form potential
is dual to an integration constant while the space-filling brane couples to a $D$-form potential that is not dual to anything at all.

In recent years it has been realized that maximal supergravities can be extended with potentials of rank $D-1$ and $D$ occurring in specific U-duality representations \cite{Riccioni:2007au,Bergshoeff:2007qi,deWit:2008ta}. In recent work \cite{Bergshoeff:2010xc,Bergshoeff:2011zk} we have developed a criterion to see which of the high-form potentials of rank $D-2, D-1$ and $D$ couple to supersymmetric defect-branes, domain walls or space-filling branes, respectively.
%Our analysis included the other potentials that couple to the standard branes as well.
Our requirement was that a gauge-invariant Wess-Zumino (WZ) term should exist
that only involves worldvolume fields that fit into a supermultiplet. This is the minimum requirement for the construction
of a kappa-symmetric worldvolume action. The result of our analysis was  that not all potentials
couple to supersymmetric branes. Furthermore, we found that
%in the case of defect-branes, domain-walls and space-filling branes,
the supersymmetric non-standard branes do not fill complete U-duality representations. A prime example of this phenomenon are the 8-forms of IIB supergravity.
They transform as the ${\bf 3}$ of ${\rm SL}(2,\mathbb{R})$ S-duality but there is only a two-dimensional space of supersymmetric configurations, spanned by the D7-brane and its S-dual \cite{Bergshoeff:2002mb}.\,\footnote{Note that there is %eric1
 a single constraint on the 9-form curvatures. This constraint is needed
for the correct counting of physical degrees of freedom in IIB supergravity but not for the correct counting of supersymmetric branes.}

In this letter we consider a specific class of branes suggested by supergravity, i.e.~the supersymmetric solitons.
These are branes whose tension scales with the inverse squared of the string coupling constant in the string frame. They are the duals
of the fundamental branes whose tension is independent of the string coupling constant.
It is well-known that a T-duality covariant formulation of the fundamental branes of toroidally compactified string theory with 32 supercharges requires a doubled geometry \cite{Hull:2004in}.
 In this letter we will probe this doubled geometry with the dual solitons.
Before discussing doubled geometry and  solitons we first review in the next section the relation between doubled geometry and fundamental branes and D-branes. For fundamental branes, doubled geometry requires an effective wrapping rule that gives rise to all the fundamental branes in a given dimension starting from ten dimensions.
We will then discuss solitonic branes.
We will first consider the standard solitons, i.e.~those with more than two transverse directions.
Probing the doubled geometry with such solitons requires an effective wrapping
rule for the solitonic branes which is dual to the wrapping rule for fundamental branes. This dual wrapping rule can be understood by the presence
of Kaluza-Klein (KK) monopoles.

We next extend our analysis to the supersymmetric non-standard
solitonic branes with less than or equal to two transverse directions whose existence is suggested by
 supergravity. These brane configurations
have been classified in our previous work  using the criterion mentioned above \cite{Bergshoeff:2011zk}.
In this letter we will show that these non-standard  solitons are
precisely obtained by applying the same dual wrapping rule as in the case of the standard solitons.
The fact that this wrapping rule works for the non-standard solitons as well cannot be
explained by the presence of the standard KK monopole alone. Instead, it
suggests the existence of a class of generalized KK monopoles in ten-dimensional string theory.


\section{Doubled Geometry}
The only fundamental brane in ten dimensions is the fundamental string. It  couples to the  background metric $g_{\mu\nu}$ via a
Nambu-Goto term and to the  NS-NS two-form potential $B_{\mu\nu}$ via a WZ term. Schematically, we have
\begin{eqnarray}
 {\cal L}^{\rm D=10}({\rm Fundamental\ String}) \ = \  T\,\sqrt{-g}\ +\ B_2\,,
\end{eqnarray}
where $T$ is the string tension. The first term at the r.h.s.~is the Nambu-Goto term containing the determinant of the pull-back of $g_{\mu\nu}$.
The second term, where we have used form notation, is the WZ term containing the pull-back of $B_{\mu\nu}$.
The special thing about fundamental branes is that
their brane tension $T$ is independent of the string coupling constant $g_s = <e^\phi>$ with $\phi$ being the dilaton.
In general the tension $T$ of a brane may scale like
$T \ \sim \ e^{\alpha\,\phi}$
in terms of an integer number $\alpha \le 0$. This leads to a classification of branes according to $\alpha$:\,\footnote{We do not consider instantons.}
\begin{eqnarray}
&&\alpha=0: \hskip .3truecm {\rm Fundamental\ Branes}\,, \nonumber \\ [.05truecm]
&&\alpha=-1: \hskip .05truecm {\rm D-branes}\,, \\ [.05truecm]
&&\alpha=-2: \hskip .1truecm {\rm Solitonic\ Branes}\,, \ \dots \nonumber \hskip 1truecm {\rm etc.}
\end{eqnarray}
Another way of classifying branes is according to the number of transverse directions. As already mentioned in the introduction we will call branes with more than two transverse directions ``standard'' and branes with less than or equal to two transverse directions ``non-standard''. Amongst the non-standard branes
we will adapt the following nomenclature:
\begin{eqnarray}
&& \#\ {\rm transverse\ directions} =2\,:\hskip .5truecm {\rm defect-branes}\,,\nonumber\\[.05truecm]
&& \#\ {\rm transverse\  directions} =1\,:\hskip .5truecm {\rm domain-walls}\,,\\[.05truecm]
&& \#\ {\rm transverse\ directions} =0\,:\hskip .5truecm {\rm space-filling\ branes}\,.\nonumber
\end{eqnarray}

%eric2
Restricting ourselves first to fundamental branes
we not only have the fundamental string in 
$D < 10$ dimensions but also  fundamental 0-branes, i.e.~wrapped strings.
They can %eric3
be attached to the fundamental string and the
corresponding WZ term gets accordingly modified with extra world-volume scalars that satisfy a self-duality condition \cite{Hull:2004in}:
\begin{equation}\label{WZstring}
  {\cal L}^{{\rm D}\le {\rm 10}}_{\rm WZ}{\rm (Fundamental\ String)} =  B_2 + \eta^{AB} {\cal F}_{1, A} B_{1 ,B}\,.
  \end{equation}
Here $B_{1,A}$ are the NS-NS 1-forms and  ${\cal F}_{1,A}=db_{0,A}+B_{1,A}$ are the 1-form world-volume curvatures of the
extra scalars $b_{0,A}$. Both  transform as a vector, indicated by the index $A$, under the T-duality group ${\rm SO}(d,d)$ with $d=10-D$.
The number of extra scalars is twice the number of compactified dimensions in line with doubled geometry \cite{Hull:2004in}. Due to
the self-duality condition that these scalars satisfy  we obtain  $(D-2) + 1/2 \cdot 2(10-D) = 8$ worldvolume degrees of freedom, where $D-2$ is the number of transverse scalars. This is the
 correct number that fits into a scalar supermultiplet.
The WZ term for the fundamental 0-branes themselves
does not contain extra scalars and is given by (omitting the explicit vector-index $A$)

\begin{equation}\label{WZ0branes}
{\cal L}^{{\rm D} \le {\rm 10}}_{\rm {WZ}} ({\rm Fundamental\ 0-Branes}) = B_1\,.
\end{equation}
In summary, in $D$ dimensions we have a T-duality vector of fundamental 0-branes and a singlet fundamental string.

Alternatively, the above counting of branes is obtained by applying the following wrapping rule for fundamental branes:
  \begin{eqnarray}\label{fundamentalwrapping}
 & & {\rm wrapped} \ \ \ \ \rightarrow\  \ \ {\rm doubled}\,, \nonumber \\
& & {\rm unwrapped} \ \ \rightarrow \ \ {\rm undoubled}\,.
 \end{eqnarray}
The doubling of branes under wrapping is due to the fact that in each dimension there is an extra fundamental 0-brane resulting from the reduction of a pp-wave. This is precisely the manifestation of T-duality. Starting from a single fundamental string in ten dimensions (either IIA or IIB)
one obtains the correct number of fundamental branes in each dimension by applying the fundamental
wrapping rule (\ref{fundamentalwrapping}) for each compactified dimension, see Table \ref{fundamental}.

\begin{table}[h]
\begin{center}
\begin{tabular}{|c||c|c|c|c|c|c|c|c|}
\hline \rule[-1mm]{0mm}{6mm} F$p$-brane &IIA/IIB& 9 & 8 & 7 & 6&5&4&3\\
\hline \hline \rule[-1mm]{0mm}{6mm} 0&&2&4&6&8&10&12&14\\
\hline \rule[-1mm]{0mm}{6mm} 1&1/1&1&1&1&1&1&1&1\\
\hline
\end{tabular}
\caption{\sl Upon applying the fundamental wrapping rule (\ref{fundamentalwrapping}) one obtains in each dimension
a singlet fundamental string and a T-duality vector of fundamental 0-branes.\label{fundamental}
}
\end{center}
\end{table}

We next consider the D-branes.
In $D=10$ dimensions fundamental strings  can end on D-branes and, accordingly, the WZ term gets
deformed by an extra Born-Infeld worldvolume vector $b_1$, with 2-form curvature ${\cal F}_2=db_1+B_2$:
\begin{equation}\label{WZterm}
  {\cal L}^{\rm {D=10}}_{{\rm WZ}}{\rm (D-branes)} = e^{{\cal F}_2}C\,.
  \end{equation}
  Here  $C$ stands for the formal sum of all RR potentials which are of odd rank for IIA and of even rank for IIB string theory.
  In \cite{Bergshoeff:2010xc} we derived the T-duality-covariant expression of the D-brane WZ terms in $D < 10$ dimensions.
Since now both wrapped and un-wrapped fundamental strings can end on the
D-branes we get a further deformation by the extra worldvolume scalars $b_{0,A}$ \cite{Bergshoeff:2010xc}:
 \begin{equation}\label{WZterm2}
  {\cal L}^{{\rm D}\le {\rm 10}}_{\rm {WZ}}{\rm (D-branes)} = e^{{\cal F}_2}e^{{\cal F}_{1,A}\Gamma^A}C\,,
  \end{equation}
where $\Gamma^A$ are the gamma-matrices of ${\rm SO}(d,d)$. The reason for the existence of the general expression (\ref{WZterm2}) is that in any dimension the
fundamental potentials transform as a singlet (2-form) and vector (1-form) under T-duality while the D-brane potentials
transform as (chiral) spinor representations of the same duality group. We have omitted these spinor indices in eq.~(\ref{WZterm2}).

At first sight the WZ term (\ref{WZterm2}) does not seem to lead to the correct counting of worldvolume degrees of freedom.
For any D$p$-brane the Born-Infeld vector
corresponds to $p-1$ degrees of
freedom. Considering also the $D-p-1$ embedding scalars one needs
only $d$ extra scalars to fill the bosonic sector of a
vector multiplet in $p+1$ dimensions. Instead, there are $2d$ scalar
fields $b_{0,A}$, that is twice too many. Unlike in the case of the fundamental string one can this time not rescue the situation by imposing a self-duality condition
on the extra scalars. Luckily, it turns out that the above counting is too naive. The expression (\ref{WZterm2}) stands for the WZ term for a whole
spinor representation of D-branes and it is enough to show that a single spinor component representing the WZ term of a particular D-brane contains only half of the $2d$ extra scalars. To show this, it is enough to expand (\ref{WZterm2}) and
consider only the first ${\cal F}_{1,A}$ term. For a given $p$ we obtain
  \begin{equation}
  C_{p+1,\alpha} + {\cal F}_{1,A} (\Gamma^A)_\alpha{}^\beta C_{p,
  \beta} + ... \quad ,
  \end{equation}
where $\alpha$ is an ${\rm SO}(d,d)$ spinor index.
Using an ${\rm SO}(d,d)$ lightcone basis, where the light-cone directions are denoted as
$A=(1\pm$, $2\pm$,..., $d\pm)$, one can show that for a given
value of the spinor index $\alpha$,  for any fixed
$n=1,...,d$, only one of the two matrices $(\Gamma^{n\pm})_\alpha{}^\beta$ gives a non-zero result when acting on a chiral
spinor. The detailed proof can be found in \cite{Bergshoeff:2011zk}.
This shows that for any
given D$p$-brane only half of the $2d$ extra scalars $b_{0,A}$ actually
occur, and this results in the correct number of degrees of freedom
for a $(p+1)$-dimensional worldvolume vector multiplet.

In summary, in each  dimension $D<10$ we have a T-duality spinor of D-branes of dimension $2^{d-1}$. These D-branes can be obtained
by applying the following D-brane wrapping rule:
  \begin{eqnarray}\label{Dbranewrapping}
 & & {\rm wrapped} \ \ \ \ \rightarrow\  \ \ {\rm undoubled}\,, \nonumber \\
& & {\rm unwrapped} \ \ \rightarrow \ \ {\rm undoubled}\,.
 \end{eqnarray}
Starting from the D-branes of ten-dimensional IIA or IIB string theory and using this wrapping rule one obtains the correct number of D-branes in $D<10$ dimensions, see Table \ref{Dbranetable}.

\begin{table}[h]
\begin{center}
\begin{tabular}{|c||c|c|c|c|c|c|c|c|}
\hline \rule[-1mm]{0mm}{6mm} D$p$-brane &IIA/IIB& 9 & 8 & 7 & 6&5&4&3\\
\hline \hline \rule[-1mm]{0mm}{6mm} 0&1/0&1&2&4&8&16&32&64\\
\hline \rule[-1mm]{0mm}{6mm} 1&0/1&1&2&4&8&16&32&64\\
\hline \rule[-1mm]{0mm}{6mm} 2&1/0&1&2&4&8&16&32&64\\
\hline \rule[-1mm]{0mm}{6mm} $\vdots$&$\vdots$&$\vdots$&$\vdots$&$\vdots$&$\vdots$&$\vdots$&$\vdots$&\\
\hline \rule[-1mm]{0mm}{6mm} 8&1/0&1&&&&&&\\
\hline \rule[-1mm]{0mm}{6mm} 9&0/1&&&&&&&\\
\hline
\end{tabular}
\caption{\sl Upon applying the D-brane wrapping rule (\ref{Dbranewrapping}) one obtains in each dimension
a T-duality spinor of D-branes. \label{Dbranetable}}
\end{center}
\end{table}

Unlike the fundamental wrapping rule the D-brane wrapping rule is self-contained, i.e.~it does not need the assistance of gravitational
solutions such as the pp-wave. The D-brane sector is also closed under duality in the sense that the dual of a D-brane is again a D-brane.
This is not the case for fundamental branes which are dual to solitonic branes. Finally, we note that all potentials 
%eric4
that couple to the supersymmetric fundamental branes and D-branes occur in the decomposition
\begin{equation}\label{decomposition}
{\rm U-duality}\ \supset\ {\rm SO}(d,d)\times \mathbb{R}^+
\end{equation}
of the U-duality representations according to which the potentials of maximal supergravity transform. The type of brane, i.e.~the value
of $\alpha$ in the tension $T=(g_s)^\alpha$, is determined by the weight of the potential under the
$\mathbb{R^+}$-scaling symmetry.
In particular, we find that the U-duality representations of the high-form potentials of rank $D-2, D-1$ and $D$
under the decomposition (\ref{decomposition}) give rise to spinor representations of T-duality corresponding to D-branes with less than or equal to two transverse directions.

This concludes our discussion of how fundamental branes and D-branes probe the doubled geometry structure. Requiring that for each brane
the corresponding dual brane also belongs to string theory it is natural to include solitons in our discussion since they are dual to the fundamental branes.
In the next section we will therefore extend our analysis to string solitons.

\section{Solitons and Dual Doubled Geometry}

We first restrict ourselves to solitonic branes with more than two transverse directions. Requiring that for
every fundamental brane there is a dual solitonic brane one finds that the following dual wrapping rule
must be introduced:
 \begin{eqnarray}\label{solitonicwrapping}
 & & {\rm wrapped} \ \ \ \ \rightarrow\  \ \ {\rm undoubled}\,, \nonumber \\
& & {\rm unwrapped} \ \ \rightarrow \ \ {\rm doubled}\,.
 \end{eqnarray}
The doubling of branes when unwrapped is due to the fact that in each dimension there is an extra
solitonic $(D-4)$-brane resulting from the reduction of a KK monopole.
Starting from the standard NS-NS five-brane of ten-dimensional IIA or IIB string theory and using the dual wrapping rule
(\ref{solitonicwrapping}) one obtains
a singlet solitonic $(D-5)$-brane and a T-duality vector of solitonic $(D-4)$-branes in each dimension $D<10$, see Table \ref{sol1table}.

\begin{table}[h]
\begin{center}
\begin{tabular}{|c||c|c|c|c|c|c|c|c|}
\hline \rule[-1mm]{0mm}{6mm} S$p$-brane &IIA/IIB& 9 & 8 & 7 & 6&5&4&3\\
\hline \hline \rule[-1mm]{0mm}{6mm} 0&&&&&&1&12&\\
\hline \rule[-1mm]{0mm}{6mm} 1&&&&&1&10&&\\
 \hline \rule[-1mm]{0mm}{6mm} 2&&&&1&8&&&\\
 \hline \rule[-1mm]{0mm}{6mm} 3&&&1&6&&&&\\
 \hline \rule[-1mm]{0mm}{6mm} 4&&1&4&&&&&\\
 \hline \rule[-1mm]{0mm}{6mm} 5&$1^\prime/1$&$1^\prime + 1$&&&&&&\\
\hline
\end{tabular}
\caption{\sl Upon applying the dual wrapping rule (\ref{solitonicwrapping}) and restricting to solitonic branes with more than two transverse directions one obtains in each dimension
a singlet solitonic $(D-5)$-brane and a T-duality vector of solitonic $(D-4)$-branes. The prime indicates that the corresponding 5-brane has a six-dimensional worldvolume tensor multiplet. \label{sol1table}
}\end{center}
  \end{table}
Unlike in the case of fundamental branes we find that the class of solitonic branes, like in the case of  D-branes,  extends to include non-standard branes, i.e.~branes with less than or
equal to two transverse directions, as well. Under the decomposition (\ref{decomposition}) of the supergravity fields the string solitons (or S-branes) organize themselves as anti-symmetric tensor representations of the T-duality group, so that in $D$ dimensions an S$p$-brane has $(p+d-5)$ indices, see Table \ref{sol2table} \cite{Bergshoeff:2011zk}.
In each dimension it is understood that the number $m$ of anti-symmetric
indices runs from $m=0$ to $m=d$, which means that the highest possible value of $p$ is 5. Furthermore, the representation with the maximum number of  indices - that is $d$ indices - splits into a self-dual and an anti-selfdual representation of the T-duality group. The self-dual representation describes solitonic five-branes with a worldvolume vector multiplet while the anti-selfdual representation involves solitonic five-branes with a worldvolume tensor multiplet.

\begin{table}[h]
\begin{center}
\begin{tabular}{|c|}
\hline \rule[-1mm]{0mm}{6mm} S$(D-5)$-brane \\
\hline  \rule[-1mm]{0mm}{6mm} [S$(D-4)$-brane]$_A$\\
\hline \rule[-1mm]{0mm}{6mm} [S$(D-3)$-brane]$_{AB}$\\
 \hline \rule[-1mm]{0mm}{6mm} [S$(D-2)$-brane]$_{ABC}$\\
 \hline \rule[-1mm]{0mm}{6mm} [S$(D-1)$-brane]$_{ABCD}$\\
 \hline
\end{tabular}
\caption{\sl String solitons transform as anti-symmetric tensor representations of the T-duality group. The number $m$ of
antisymmetric indices runs from $m=0$ to $m=d$. \label{sol2table}}
\end{center}
\end{table}

Unlike the singlet and vector solitons, it turns out that not all components of the higher-rank antisymmetric tensor representations
correspond to %eric5
{\it supersymmetric} solitons. To understand this it is enough to consider only the leading and subleading term in the
 WZ term of a solitonic S$p$-brane which takes the schematic form ($m = p+d-5$) \cite{Bergshoeff:2011zk}
\begin{equation}
D_{A_1\dots A_m} + {\overline {{\cal G}(c)}}\,\Gamma_{A_1\dots A_m}\,C\  + \dots \,.
\end{equation}
Here $D$ is the solitonic target space potential, $C$ is the formal sum of RR target space gauge fields and ${\cal G}(c)$ is the formal sum of worldvolume RR $n$-form gauge fields $c$. Each $n$-form represents a possible D-brane ending on the soliton.
Note that both the target space potentials $C$ and the worldvolume potentials $c$ transform as spinors under T-duality.
In general there are too many worldvolume potentials to fit a supermultiplet.
We therefore need to limit the number of worldvolume gauge fields as much as possible such that the independent ones do fit into a supermultiplet. One way to restrict this number is by imposing worldvolume duality conditions. It turns out that this is not enough \cite{Bergshoeff:2011zk}.
To obtain a supersymmetric soliton one
also needs to restrict the number of T-duality directions in the anti-symmetric tensor representation such that the $\Gamma_{A_1\dots A_m}$ matrix projects out the correct number of worldvolume gauge fields.
Using a lightcone basis $A=(1\pm\,, 2\pm\,, \dots  ,d\pm)$ we found that only the
 components
\begin{equation}\label{countingrule}
[ABC \dots ]= [m\pm n\pm p\pm \dots]\ \ \ {\rm with}\ \ \ m\ne n \ne p \dots
\end{equation}
are supersymmetric. For more details, we refer to  \cite{Bergshoeff:2011zk}. This leads to a precise prediction of the number of supersymmetric solitons in each dimension. For instance, in $D=6$ dimensions there are solitonic
domain-walls  transforming in the
three-index anti-symmetric tensor representation ${\bf 56}$ of the ${\rm SO(4,4)}$ T-duality group.
This representation occurs in the decomposition (\ref{decomposition}) of a 5-form gauge potential that transforms in the ${\bf 144}$ of the ${\rm SO}(5,5)$ U-duality group.
According to the counting rule (\ref{countingrule}) only the directions $1\pm 2\pm 3\pm\,, 1\pm 2\pm 4\pm\,, 1\pm 3\pm 4\pm$ and $2\pm3\pm4\pm$ correspond to supersymmetric solitons. We therefore find that only 32 out of
the ${\bf 56}$ configurations are supersymmetric.

Remarkably, precisely the same numbers of supersymmetric solitons are obtained by simply extending the dual wrapping rule (\ref{solitonicwrapping})
to the non-standard solitons as well, see Table \ref{sol3table}. This includes the 32 supersymmetric
solitonic domain-walls in $D=6$ dimensions mentioned in the example above.
This extension is non-trivial in the sense that not only the singlet soliton doubles when unwrapped but the other solitons double as well when unwrapped. Whereas the doubling of the singlet soliton can be understood by the presence of a singlet KK monopole in each dimension, a similar explanation for the doubling of the other solitons is not available.
We will discuss a possible interpretation of this result
in the next section.

\begin{table}[h]
\begin{center}
\begin{tabular}{|c||c|c|c|c|c|c|c|c|}
\hline \rule[-1mm]{0mm}{6mm} S$p$-brane &IIA/IIB& 9 & 8 & 7 & 6&5&4&3\\
\hline \hline \rule[-1mm]{0mm}{6mm} 0&&&&&&1&12&84\\
\hline \rule[-1mm]{0mm}{6mm} 1&&&&&1&10&60&280\\
 \hline \rule[-1mm]{0mm}{6mm} 2&&&&1&8&40&160&560\\
 \hline \rule[-1mm]{0mm}{6mm} 3&&&1&6&24&80&240&\\
 \hline \rule[-1mm]{0mm}{6mm} 4&&1&4&12&32&80&&\\
 \hline \rule[-1mm]{0mm}{6mm} 5&$1^\prime/1$&$1^\prime + 1$&$2^\prime + 2$&$4^\prime + 4$&$8^\prime + 8$&&&\\
\hline
\end{tabular}
\caption{\sl By applying the dual wrapping rule (\ref{solitonicwrapping}) to both standard and non-standard solitons one obtains precisely the number of supersymmetric solitons predicted by the counting rule (\ref{countingrule}).
 The prime indicates  5-branes with a six-dimensional worldvolume tensor multiplet. \label{sol3table}}
\end{center}
\end{table}

\section{Generalized Kaluza-Klein monopoles}

To understand the wrapping rule for standard solitons it is enough to consider the standard KK monopole only. In $D=10$ dimensions the monopole solution is
characterized by 6 worldvolume, one isometry and three transverse directions. To obtain a brane one must reduce over the
isometry direction which leads to a solitonic S5-brane. The KK monopole is magnetically charged with respect to the KK vector which
is represented by off-diagonal components of the metric.
Formally, one may therefore say that the KK monopole is electrically charged with respect to the dual graviton.
Although a dual graviton mixed-symmetry tensor  $D_{7,1}$ can only be defined at the linearized level, see e.g.~\cite{Bergshoeff:2008vc},
for the present purposes it is convenient to introduce such a potential as an organizing principle. Upon reduction to $D=9$
dimensions a mixed-symmetry field $D_{7,1}$ gives rise to both a 7-form and a 6-form potential. Only the 6-form potential
is dual to the KK vector and corresponds to the KK monopole. The 7-form potential is dual to the KK scalar and does not correspond to
a supersymmetric soliton. We therefore need to restrict the possible reductions of $D_{7,1}$ with the condition that when the index
after the comma in $7,1$ is internal also one of the indices before the comma has to be internal. The reduction to $D=9$ dimensions of the NS-NS
solitonic 5-brane together with the $D=10$ KK monopole, represented by the mixed-symmetry tensor $D_{7,1}$,  then leads to the desired dual wrapping rule (\ref{solitonicwrapping})
(here $\sharp$ denotes the internal direction)
 \begin{eqnarray}
  & & D_6 \rightarrow D_{5 \sharp}\,, \quad D_6\,, \nonumber\\
  & & D_{7,1} \rightarrow D_{6 \sharp, \sharp}\,.
 \end{eqnarray}
This yields a singlet S4-brane and an SO(1,1) vector of S5-branes, one with a vector and one with a tensor multiplet.
This works for any dimension. For instance, reducing to $D=7$ dimensions we obtain
 \begin{eqnarray}
  & & D_6 \rightarrow D_{3 ijk} (1)\quad D_{4 ij} (3)\,,
  \nonumber\\
  & & D_{7,1} \rightarrow D_{4 ijk, i}\,, (3)\,,
 \end{eqnarray}
where $i=1,2,3$ is a $\rm{GL}(3)$ index. The number between brackets indicates the number of potentials. Again we obtain a singlet S2-brane
and an SO(3,3) T-duality vector of S3-branes.

We next extend the analysis to include the non-standard solitons. For the dual wrapping rule to work in this case as well,
we need an extra inflow of branes from objects which  we call generalised KK monopoles. We represent these
extra objects by mixed-symmetry fields, that a priori can be of the generic form $D_{m,n,p,...}$ for $m\ge n \ge p $ non-negative integers that denote the number of separately antisymmetric indices. Surprisingly, the following set of fields suffices:
 \begin{equation}\label{mixedsymmetry}
  D_{6+n,n} \quad \quad ,\quad \quad n=0,1,2,3,4 \quad ,
  \end{equation}
where  $6+n,n$ 
refers to the symmetries corresponding to a Young tableau with two columns, 
one with $6+n$ entries and a second one with $n$ entries.
This includes the fields $D_6$ (the dual NS-NS 2-form), $D_{7,1}$ (the dual
graviton), $D_{8,2}$ (which is another dual to the NS-NS 2-form) together with the higher-rank fields  $D_{9,3}$ and
$D_{10,4}$. The rule for all fields is that they give rise to
supersymmetric branes only when the $n$ indices on the right of the comma in
$D_{6+n,n}$ are
compactified along directions on which also $n$ of the $6+n$ indices on the left of the comma
 are compactified. This reduction rule implies that there are
no solitonic branes with a  worldvolume dimension higher than 6.

The restricted reduction rule for the mixed-symmetry fields (\ref{mixedsymmetry}) yields exactly the right number of additional solitons such that the dual wrappping rule (\ref{solitonicwrapping})
works. For instance, the reduction to $D=6$ dimensions yields the following potentials ($i=1,2,3,4$ is a GL(4) index)\,:
  \begin{eqnarray}
  & & D_6 \rightarrow D_{2 ijkl} (1)\quad D_{3 ijk} (4) \quad
  D_{4ij}
  (6) \quad D_{5i} (4) \quad D_6 (1)
  \nonumber\\
  & & D_{7,1} \rightarrow D_{3 ijkl, i} (4)\quad D_{4ijk,i} (12) \quad D_{5ij,i}(12) \quad D_{6i,i} (4)\nonumber
  \\
  & & D_{8,2} \rightarrow D_{4ijkl,ij} (6) \quad D_{5ijk,ij} (12)
  \quad D_{6ij,ij} (6)
  \nonumber \\
  & & D_{9,3} \rightarrow D_{5ijkl,ijk} (4) \quad D_{6ijk,ijk}(4)
  \nonumber \\
   && D_{10,4} \rightarrow D_{6ijkl,ijkl} (1) \quad .
  \end{eqnarray}
  This yields precisely the sequence of 1,8,24,32,16   potentials that can  be found in the $D=6$ column of Table \ref{sol3table}.
  The other dimensions work in the same way.

It is not clear what the precise status of the mixed-symmetry fields (\ref{mixedsymmetry}) for $n=2,3,4$ is. One point of view is to consider these fields
as a formal framework to get a handle on the  properties of the non-standard solitons after reduction. According to this point of view
they should not themselves be associated with objects in ten-dimensional string theory. A more exciting possibility is that the mixed-symmetry potentials $D_{6+n,n}$ for $n=2,3,4$ can be associated with non-standard KK monopoles in string theory in the same way
that the mixed-symmetry tensor $D_{7,1}$ encodes information about the standard KK monopole. It is suggestive
to conjecture that the mixed-symmetry potentials $D_{6+n,n}$ represent supersymmetric solutions with 6 worldvolume directions, $n$ isometry directions and $4-n$ transverse directions. For $n=0$ this is the NS-NS 5-brane and for
$n=1$ this is the standard KK monopole. The $n=2,3,4$ cases can probably be represented as supersymmetric
single brane solutions by uplifting the lower-dimensional solitons. However, since they have less than or equal to 2 transverse directions they will  not be well-defined by themselves. For instance, in the case of 2 transverse
directions it is likely that one should consider multiple brane configurations to obtain finite energy solutions.
This requires a further investigation.

\section{Conclusions}

In this letter we first mentioned that the wrapping rule for fundamental branes, to be consistent with T-duality, requires
extra 0-branes originating from the reduction of the pp-wave. The D-branes are consistent by themselves and their wrapping rule does not require
any additional object.  The fundamental branes are mapped under duality to the standard solitons.
Accordingly, the wrapping rule for these standard solitons  requires the dual of the pp-wave which is the KK-monopole.

We next extended the
analysis to the supersymmetric non-standard solitons which have been classified using supergravity input. Remarkably, these non-standard solitons result from the same dual wrapping rule that leads to the standard solitons. This rule can, however, not be explained by another use
of the KK monopole. Other objects are needed and we showed that the corresponding fields are a limited number of mixed-symmetry tensors given in
(\ref{mixedsymmetry}). Remarkably, these solitonic mixed-symmetry fields are all contained in the solitonic sector of the spectrum of the very-extended Kac-Moody algebra ${\rm E}_{11}$ \cite{West:2001as}. It will be interesting to see whether this ${\rm E}_{11}$ algebra
can play a guiding role in understanding the organizing principle here.

One may extend the present analysis by considering branes whose tension scales as $(g_s)^\alpha$ with $\alpha \le -3$. The first objects to consider are the ones with $\alpha=-3$. In ten-dimensional string theory there is only one such object  and that is the S-dual of the D7-brane. We will call these exceptional branes E-branes. It turns out that E-branes occur in any dimension as tensor-spinor representations of the T-duality group \cite{inpreparation}. The supersymmetric ones can be classified by considering their WZ term. To obtain the
lower-dimensional supersymmetric E-branes from ten dimensions one needs
the following exotic wrapping rule \cite{inpreparation}
 \begin{eqnarray}\label{exoticwrapping}
 & & {\rm wrapped} \ \ \ \ \rightarrow\  \ \ {\rm doubled}\,, \nonumber \\
& & {\rm unwrapped} \ \ \rightarrow \ \ {\rm doubled}\,.
 \end{eqnarray}
To realize this wrapping rule one needs a further inflow of branes resulting from other objects in string theory.
%eric5
A similar wrapping rule explaining the occurrence of branes in lower dimensions with $\alpha \le -4$ does not seem to exist \cite{inpreparation}.

%eric6
In summary, the classification of supersymmetric branes in lower dimensions with  $0 \le  \alpha \le -3$, suggested by supergravity, can be reproduced by a set of simple wrapping rules.
%eric7
These rules suggest the existence of a set of  additional objects in ten-dimensional string theory. It
remains to be  seen wether these objects indeed  have a meaning within string theory.

\section*{Acknowledgements}

E.B. wishes to thank the hospitality of King's College London, where part of this work was done. F.R. wishes to thank the University of Groningen for hospitality.

%% References


\bibliographystyle{elsarticle-num}
% \bibliography{<your-bib-database>}



\end{document}